# Concentration retrieval in a calibration-free wavelength modulation spectroscopy system using particle swarm optimization algorithm

TINGTING ZHANG,[1,3] YONGJIE SUN,[2,3] PENGPENG WANG,[1] AND CUNGUANG ZHU[1,*]

[1] *School of Physical Science and Information Technology, Liaocheng University, Liaocheng 252000, China*
[2] *School of Physical Science and Technology, University of Jinan, Jinan 250022, China*
[3] *Contributed equally*
*\*Corresponding author: zhucunguang@lcu.edu.cn*

**Abstract:** This paper develops a spectral fitting technology based on the particle swarm optimization (PSO) algorithm, which is applied to a calibration-free wavelength modulation spectroscopy system to achieve concentration retrieval. As compared with other spectral fitting technology based on the Levenberg-Marquardt (LM) algorithm, this technology is relatively weakly dependent on the pre-characterization of the laser parameters. The gas concentration is calculated by fitting the simulated spectra to the measured spectra using the PSO algorithm. We validated the simulation with the LM algorithm and PSO algorithm for the target gas $C_2H_2$. The results showed that the convergence speed of the spectral fitting technique based on the PSO algorithm was about 63 times faster than the LM algorithm when the fitting accuracy remained the same. Within 5 seconds, the PSO algorithm can produce findings that are generally consistent with the values anticipated.

## 1. Introduction

Wavelength modulation spectroscopy (WMS) technology has been extensively used to measure gas characteristics in areas such as greenhouse gas and atmospheric pollutant monitoring [1,2], combustion monitoring [3-5], and industrial process control [6-8] due to its noise rejection, high measurement accuracy, simplicity and fast response time [9-13].

In early applications, WMS requires calibration with a gas mixture of known concentrations, which limits the utilization of WMS technology in harsh environments (such as high temperature or pressure) or where the relevant gas conditions are poorly known. Thus, researchers have developed numerous calibration-free techniques to address this issue [14-21]. Li *et al*. [22] proposed a calibration-free technique combining laser tuning parameters with a WMS analytical model of absorption spectrum, which can eliminate the majority of calibration factors and infer gas parameters. Rieker *et al*. [23] utilized the first harmonics to normalize the second harmonics (WMS-2*f*/1*f*) and extended the WMS analysis model to the field of combustion monitoring to retrieve the temperature and concentration of water vapor simultaneously. Yang *et al*. [24] proposed a gas concentration retrieval approach based on the first harmonic phase angle ($\theta_{1f}$) method, which is immune to the laser intensity and the demodulation phase. A second harmonic phase angle method ($\theta_{2f}$) based on WMS was proposed for trace gas detection enabling background-free detection and immunity to light intensity fluctuations. [25] In the past decades of calibration-free WMS method development, Levenberg-Marquardt (LM) algorithm has been widely used to deal with nonlinear least-squares problems in spectral fitting for gas concentration retrieval. Christopher *et al.* [26] proposed a strategy that enables accurate calibration-free WMS measurements of gas properties without needing prior knowledge of the transition line-shape parameters. Combining the LM algorithm with dual-spectroscopy techniques, Li *et al*. [27] developed a mid-infrared laser trace gas sensor capable of online monitoring the concentration of multi-component gases. Raza *et*

*al.* [28] utilized the LM algorithm for least-squares fitting of simulated and measured scanned-WMS-2*f*/1*f* spectrum to concurrently monitor information on temperature and concentration changes of CO and NH$_3$ in high-temperature environments.

The LM algorithm has been extensively used to assess the gas conditions of calibration-free WMS systems. It is worth emphasizing that the fitting by the LM algorithm is often unsatisfactory when models with multiple free parameters are encountered. The reason is that an increase in the number of free parameters leads to a rise in the order of the LM algorithm's operation matrix, which deteriorates the LM algorithm's accuracy and fitting efficiency. The challenge can be avoided by using the methodology of pre-characterization with laser parameters in the WMS models, which decreases the number of free parameters. However, this undoubtedly adds the cost and complexity of the testing process. At the same time, it may cause a series of secondary problems, such as measurement errors caused by the failure of the characterized values during the long operation of the instrument.

In this paper, we propose a spectral fitting technology based on the particle swarm optimization (PSO) algorithm, which is applied to a calibration-free wavelength modulation spectroscopy system to achieve concentration retrieval. Contrasted with the spectral fitting technology based on the LM algorithm, this technology is relatively weakly dependent on the pre-characterization of the laser parameters. For the target gas C$_2$H$_2$, we evaluate the LM algorithm and PSO algorithm comparatively by simulation. The findings demonstrated that, notably in the multi-parameter model without exact characterization, the spectral fitting technique based on the PSO algorithm performs better in terms of convergence speed and fitting accuracy.

## 2. Theory and methodology

The WMS-2*f*/1*f* is the most representative calibration-free system. In this study, we will evaluate the performance of the PSO algorithm based on the WMS-2*f*/1*f* system, and in this section, we will first briefly describe the theory of the WMS-2*f*/1*f* system, and then introduce the idea and process of the PSO algorithm to implement the spectral line fitting in this system.

### 2.1 Theory of WMS-2f/1f

In wavelength modulation spectroscopy, the output of the laser is modulated periodically by varying the current injection to produce modulated light intensity $I_0(t)$ and modulated light frequency $v(t)$:

$$I_0(t) = \overline{I_0}\left[1 + i_1\cos(\omega t + \psi_1) + i_2\cos(2\omega t + \psi_2)\right] \quad (1)$$

$$v(t) = v_c + \Delta v \cdot \cos(\omega t) \quad (2)$$

where $\overline{I_0}$ is the average light intensity at the laser modulation center frequency $v_c$, $i_1$ and $i_2$ are the intensity amplitudes of the first- and second-order (normalized by $\overline{I_0}$), modulation angular frequency $\omega = 2\pi f_m t$, $f_m$ is laser modulation frequency, $\psi_1$ and $\psi_2$ are the phase shift between frequency modulation (FM) and linear and nonlinear intensity modulation (IM), respectively, $\Delta v$ is the modulation depth of the laser frequency.

According to Beer-Lambert law, when the laser beam passes through a gas cell filled with the absorbing gas, the transmitted laser intensity at frequency $v$ can be expressed as:

$$I_t(t) = I_0(t) \cdot \exp\left[-\alpha(v(t))\right] \quad (3)$$

where $I_0(t)$ is incident laser intensity, $\alpha(v(t))$ is the spectral absorbance, in the case of weak absorption ($\alpha < 0.05$).

$$I_t(t) = I_0(t)\cdot\exp\left[-\alpha(v(t))\right] \approx I_0(t)\cdot\left[1-\alpha(v(t))\right] = I_0(t)\cdot\left[1-PS(T)CLg(v,v_0)\right] \quad (4)$$

where $I_t(t)$ is the transmitted light intensity, $P$ is the total gas pressure, $S(T)$ is the intensity of the absorption spectrum at $T$ temperature, $C$ is the concentration of the gas to be measured, $L$ is the length of the optical range, $g(v, v_0)$ is the line shape function at optical frequency $v$ of the absorption feature.

At atmospheric pressure, $g(v, v_0)$ can be expressed as Lorentzian line shape function:

$$g(v, v_0) = \frac{2}{\pi \Delta v_c} \frac{1}{1+[x+m\cos(\omega t)]^2} \tag{5}$$

$$x = 2\frac{v_c - v_0}{\Delta v_c} \tag{6}$$

$$m = \frac{2\Delta v}{\Delta v_c} \tag{7}$$

where $\Delta v_c$ is the full width at half-maximum, $v_0$ is the line-center frequency of the absorption spectrum, $x$ is the normalized frequency, $m$ is the modulation index.

when the laser is modulated by a sinusoidal injection current, $\exp[-\alpha(v(t))]$ can be expanded in a Fourier cosine series as follows:

$$\exp[-\alpha(v(t))] = \sum_{k=0}^{\infty} H_k(v_c, \Delta v) \cdot \cos(k\omega t) \tag{8}$$

The $I_t(t)$ signal is input into the lock-in amplifier (LIA) and multiplied with a reference cosine wave and sine wave at $nf$ to extract the $X_{nf}$ and $Y_{nf}$ signals, respectively, which are subsequently low-pass filtered to determine $S_{2f/1f}$, given by Eq. (9),

$$S_{2f/1f} = \sqrt{(\frac{X_{2f}}{X_{1f}})^2 + (\frac{Y_{2f}}{Y_{1f}})^2}$$

$$= \sqrt{\left(\frac{H_2 + \frac{i_1}{2}(H_1+H_3)\cos\psi_1 + i_2 H_0 \cos\psi_2}{H_1 + i_1(H_0 + \frac{H_2}{2})\cos\psi_1 + \frac{i_2}{2}(H_1+H_3)\cos\psi_2}\right)^2 + \left(\frac{\frac{i_1}{2}(H_1-H_3)\sin\psi_1 + i_2 H_0 \sin\psi_2}{i_1(H_0 - \frac{H_2}{2})\sin\psi_1 + \frac{i_2}{2}(H_1-H_3)\sin\psi_2}\right)^2} \tag{9}$$

From Eq. (9), it is clear that the spectral model $S_{2f/1f}$ is a function of multiple parameters, including the modulation index $m$, gas concentration $C$, the linear and nonlinear IM depth $i_1$ and $i_2$, and the phases shift $\psi_1$ and $\psi_2$ between FM and linear and nonlinear IM, among other parameters. This presents a significant obstacle for the LM algorithm, which we shall successfully resolve using the PSO algorithm.

### 2.2 PSO-based calibration-free WMS spectral fitting technique

The PSO algorithm is an evolutionary computing method based on swarm intelligence that mimics the social interaction and migration of bird flocks as well as the social sharing of knowledge to aid in the evolution of individuals. Its core idea is to use the information interaction among a population of particles to seek the optimal solution to a multi-parameter optimization problem. In this paper, we apply it to a multi-parameter optimization of the spectral model $S_{2f/1f}$ for gas concentration and laser parameters retrieval in a calibration-free WMS system.

In the proposed technique, the simulated spectra are fitted to the measured spectra. The concentration is obtained from the optimal solution of the fitting, which error is determined by the objective function $F(\eta_{id})$,

$$F(\eta_{id}) = \sum_{n=1}^{N}(f_i(x_n, \eta_{id}) - y_n)^2 \tag{10}$$

where $x_n$ is the normalized frequency, $n=1, 2, 3, ..., N$, $N$ is the total number of sampling points in the simulated spectra, $f_i(x_n, \eta_{id})$ is simulated spectral data given by Eq. (9), the free parameters are denoted by the vector $\eta$ that defines the current position of the particle, $D$ is dependent on the number of free parameters ($d=1, 2, 3, ..., D$), $K$ is the population of particle ($i=1, 2, 3, ..., K$), $y_n$ is the measured spectra.

The PSO-based calibration-free WMS spectral fitting technique flow chart is shown in Fig. 1. In this technique, the following parameters can all be set to free parameters of fitting, including the modulation index $m$, gas concentration $C$, the linear and nonlinear IM depth $i_1$ and $i_2$, and the phases shift $\psi_1$ and $\psi_2$ between FM and linear and nonlinear IM, respectively.

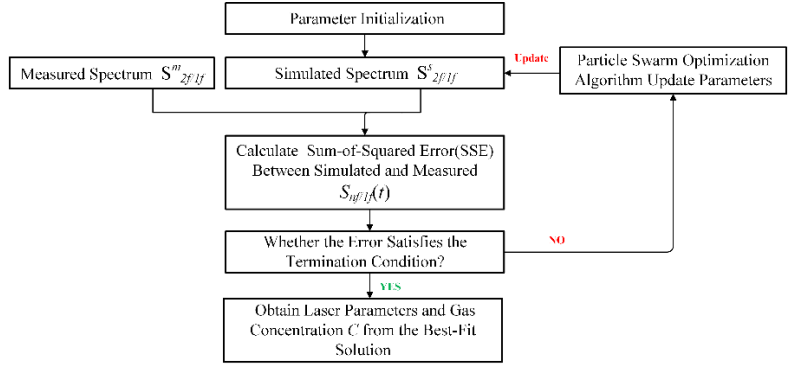

Fig. 1. Flow chart for seeking the optimal solution of gas properties and laser parameters using the PSO algorithm.

Step 1: Parameter initialization - Generate $K$ vectors $\eta_i$ for the fitting procedure, which are randomly assigned within a specific range, $i= 1, 2, …, K$.

Step 2: Acquisition of simulated spectra - The simulated spectra $f_i(x_n, \eta_{id})$ are obtained by substituting the initialized parameters $\eta_i$ into Eq. (9).

Step 3: Parameters updating - The objective function $F$ of Eq. (10) is used to update the fitness values of each particle, which measures the closeness of the corresponding solution with the optimal solution.

Step 4: Judgment - If the convergence between measured spectra $S^m_{2f/1f}$ does not satisfy the termination condition of the optimization, then continue to Step 5; Otherwise, perform Step 6.

Step 5: Particle movement - Particles rely on velocity and position updates to complete optimization, which determines the search route of particles in space. The position of the particle $i$ in the $(t+1)$th iteration is updated by the following equation:

$$x_{id}^{t+1} = x_{id}^t + v_{id}^{t+1} \qquad (11)$$

where $x^t_{id}$ is the position of the particle $i$ in the $t$th iteration, the velocity of the particle $i$ in the $(t+1)$th iteration $v^{t+1}_{id}$ can be defined as:

$$v_{id}^{t+1} = \omega \cdot v_{id}^t + c_1 rand()\left(p_{best}^t - x_{id}^t\right) + c_2 rand()(g_{best}^t - x_{id}^t) \qquad (12)$$

where $v^t_{id}$ is the velocity of the particle $i$ of $t$th iteration, $\omega \in [0,1]$ is the inertia weight, rand() $\in [0,1]$ is a random value, $p^t_{best}$ is individual best position of the particle $i$ in the $t$th iteration, $g^t_{best}$ is the global best position of the any particle in the $t$th iteration, $c_1$ and $c_2$ are the real acceleration coefficients that control how much the global and individual best positions should influence the velocity of the particle.

After updating of each particle position and fitness value in steps 3 and 5, the swarm of particles will keep moving closer to the region of the optimal solution.

Step 6: Termination condition - If the convergence between measured spectra $S^m_{2f/1f}$ and simulated spectra $S^s_{2f/1f}$ satisfies the termination condition of the optimization, the value of best-fit solution is output at this point, and laser parameters and gas concentration $C$ are obtained.

### 3. Simulation verification

The feasibility of the PSO-based WMS spectral fitting technique is verified in the following. In order to avoid the characterization errors introduced by the metrology instruments affecting the evaluation of the algorithm performance, we chose to perform the validation by simulation in MATLAB R2019b platform (the computer has the following specifications: CPU: i5-10400F, RAM: 8GB). We compared the spectral fitting effects of the LM algorithm and the PSO algorithm under the same conditions.

The P(13) spectral line of acetylene gas at 1532.83 nm was chosen as the target spectral line. Given the values of the spectral parameters in Table 1, the measured spectra that should have been collected in the experiment were replaced by the simulated data set calculated by Eq. (9):

$$y_n = [(x_1, y_1), (x_2, y_2), ..., (x_{4000}, y_{4000})] \quad (13)$$

The above virtual measured spectrum is first fitted using the LM algorithm, which is discussed in two cases. In the first case, only the gas concentration $C$ is used as a free parameter, and all other parameters are assigned to known values. The results are shown in Fig. 2. The results of fitting for the second case are shown in Fig. 3, where six parameters are set as free parameters, which are the gas concentration $C$, the linear and nonlinear IM depths $i_1$ and $i_2$, and the phases shifts $\psi_1$ and $\psi_2$ between FM and linear and nonlinear IM, respectively. In both cases, the convergence time is limited to 100 s. According to Fig. 2 and Fig. 3, when the number of free parameters goes from one to six, the residuals grow by 10 orders of magnitude ($10^{-14}$ to $10^{-4}$). The fitting effect of the LM algorithm deteriorates significantly with the increase in the number of free parameters.

**Table 1. Summary of spectral parameters**

| Symbol | Quantity | Value |
|---|---|---|
| $i_1$ | linear IM depth at line-center frequency | 0.15 |
| $i_2$ | nonlinear IM depth at line-center frequency | $3 \times 10^{-3}$ |
| $\psi_1$ | phases shift between FM and linear IM | $0.6\pi$ rad |
| $\psi_2$ | phases shift between FM and nonlinear IM | $0.5\pi$ rad |
| $m$ | modulation index | 1.5 |
| | isotopologue | $^{12}C_2H_2$ |
| $v_c$ | line-center frequency of transition | 6523.8792 cm$^{-1}$ |
| $\Delta v_c/2$ | half width at half-maximum | 0.0777 cm$^{-1}$ |
| $S$ | line strength | $1.035 \times 10^{-20}$ cm/mol |
| $T$ | arbitrary temperature | 296 K |
| $P$ | total pressure | 1 atm |
| $L$ | absorption path length | 20 cm |

To explore the fitting performance of the PSO algorithm, we use the PSO algorithm to fit the identical virtual measured spectrum $y_n$ with a free parameter $\eta_2$ as the LM algorithm. Fig. 4 shows the spectral fitting effect of the PSO algorithm with a 1.6 s convergence time. As can be seen, the PSO-based spectral fitting technique has a faster convergence rate than the LM algorithm by a factor of roughly 63 while maintaining residuals at $10^{-4}$. To better demonstrate the performance of the PSO algorithm in the multi-parameter model fitting, we increased the convergence time to 5 s, which output best-fit free parameters are shown in Table 2.

Table 3 shows the best-fit free parameters obtained from the LM algorithm fitting in Fig. 3. It is clear that the relative errors predicted by the LM algorithm for the free parameters, such as the linear and nonlinear IM depth $i_1$ and $i_2$, and the phases shift $\psi_1$ between the linear IM and FM exceed 5%. However, the fitting effect of the PSO algorithm is significantly improved compared to the LM algorithm, where the relative errors of free parameters are less than 1%. The predicted values are essentially consistent with the expected value.

In summary, the PSO algorithm outperforms the LM algorithm in terms of convergence time and error for the model with multiple free parameters. Satisfactory multi-parameter optimization can get rid of the dependence on pre-characterization and also prevent the measurement error caused by pre-characterization failure.

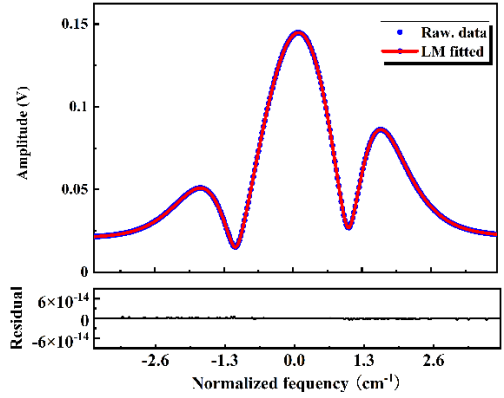

Fig. 2. Fitting effect of the LM algorithm with a convergence time of 100 s and a fitting parameter of $\eta_1=[C]$.

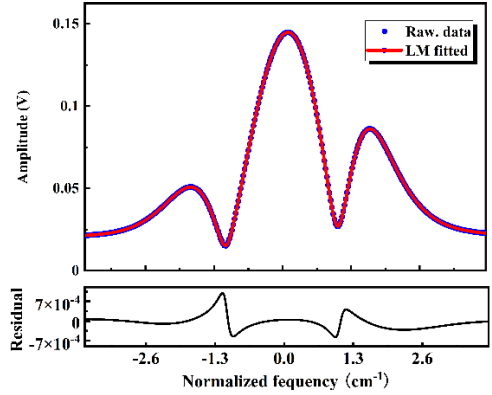

Fig. 3. Fitting effect of the LM algorithm with a convergence time of 100 s and a fitting parameter of $\eta_2=[m, C, i_1, i_2, \psi_1, \psi_2]$

Table 2. The best-fit free parameters predicted by the PSO algorithm with a convergence time of 5 s

| Free parameters | Expected value | Predicted of the PSO algorithm | Relative Errors |
| --- | --- | --- | --- |
| $m$[cm$^{-1}$] | 1.500 | 1.500 | 0.00 |
| $c$[ppmv] | 400.0 | 400.5 | 0.13% |
| $i_1$[pm/mA] | 0.150 | 0.149 | 0.67% |
| $i_2$[pm/mA] | 0.003 | 0.003 | 0.00 |
| $\varphi_1[\pi]$ | 0.600 | 0.601 | 0.17% |
| $\varphi_2[\pi]$ | 0.500 | 0.500 | 0.00 |

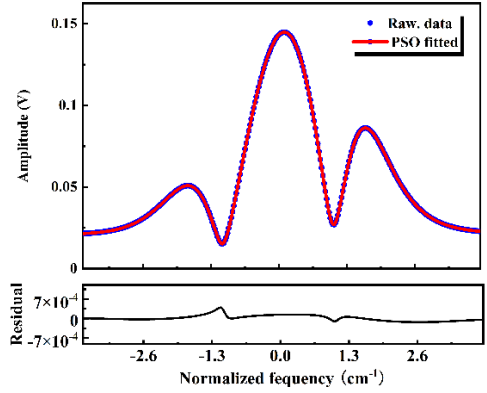

Fig. 4 Fitting effect of the PSO algorithm with a convergence time of 1.6 s and a fitting parameter of $\eta_2 = [m, C, i_1, i_2, \psi_1, \psi_2]$.

Table 3. The best-fit free parameters predicted by the LM algorithm with a convergence time of 100 s

| Free parameters | Expected value | Predicted of the LM algorithm | Relative Errors |
|---|---|---|---|
| $m$[cm$^{-1}$] | 1.500 | 1.529 | 1.93% |
| $c$[ppmv] | 400.0 | 410.7 | 2.68% |
| $i_1$[pm/mA] | 0.150 | 0.1677 | 11.8% |
| $i_2$[pm/mA] | 0.003 | 0.0034 | 13.3% |
| $\varphi_1[\pi]$ | 0.600 | 0.647 | 7.83% |
| $\varphi_2[\pi]$ | 0.500 | 0.487 | 2.60% |

## 4. Conclusion

This paper proposes a spectral fitting technology based on the PSO algorithm, which is applied to a calibration-free wavelength modulation spectroscopy system to achieve concentration retrieval. Contrasted with the spectral fitting technique based on the classical LM algorithm, the retrieval of gas concentrations by this technique is weakly dependent on the pre-characterization of the laser parameters. The simulation results indicate that the relative errors of best-fit parameters $i_1$, $i_2$, and $\psi_1$ predicted by the LM algorithm exceed 5%, but all predicted by the PSO algorithm are less than 1%. The convergence speed of the PSO algorithm was about 63 times faster than the LM algorithm when the fitting accuracy remained the same. All these results prove that the PSO algorithm outperforms the LM algorithm in terms of convergence time and error for the model with multiple free parameters.

**Funding.** National Natural Science Foundation of China (61705080); Promotive Research Fund for Excellent Young and Middle-aged Scientists of Shandong Province (ZR2016FB17, BS2015DX005); the Youth Innovation Technology Support Program of Shandong Province Colleges and Universities under Grant(2019KJJ011); Scientific Research Foundation of Liaocheng University (318012101); Startup Foundation for Advanced Talents of Liaocheng University (318052156,318052157).

**Disclosures.** The authors declare no conflicts of interest.

**Data availability.** Data underlying the results presented in this paper are not publicly available at this time but may be obtained from the authors upon reasonable request.